\documentclass[12pt]{article}

\usepackage{amssymb}
\usepackage{graphics}
\usepackage{makeidx}
\usepackage{layout}
\usepackage{amsmath}
\usepackage[vcentermath]{youngtab}

\def\be{\begin{equation}}
\def\ee{\end{equation}}
\def\bea{\begin{eqnarray}}
\def\eea{\end{eqnarray}}
\def\nn{\nonumber}

\def\de{\partial}

\def\ie{{\it i.e.}~}

\def\ap{{\alpha^\prime}}

\def\a{\alpha}
\def\b{\beta}

\def\m{\mu}
\def\n{\nu}

\def\cA{{\cal A}}
\def\cB{{\cal B}}
\def\cC{{\cal C}}

\def\cI{{\cal I}}
\def\cJ{{\cal J}}

\def\cR{{\cal R}}
\def\cS{{\cal S}}
\def\cT{{\cal T}}

\def\cV{{\cal V}}

\begin{document}
\begin{titlepage}
\begin{flushright}
{ROM2F/2011/10}
\end{flushright}
\vskip 1cm
\begin{center}
{\Large\bf Scattering higher spins
\\ \vspace{4mm} off D-branes }\\
\end{center}
\vskip 2cm
\begin{center}
{\bf Massimo Bianchi} and {\bf Paolo Teresi} \\
{\sl Dipartimento di Fisica, Universit\`a di Roma ``Tor Vergata''\\
I.N.F.N. Sezione di Roma ``Tor Vergata''\\
Via della Ricerca Scientifica, 00133 Roma, ITALY}\\
\end{center}
\vskip 1.0cm
\begin{center}
{\large \bf Abstract}
\end{center}

We study scattering amplitudes of higher spin closed-string states off D-branes. For states in the `first Regge trajectory' we find remarkable simplifications both at tree level and at one loop. We discuss the high energy behavior in the Regge limit and comment on the validity of the eikonal approximation in this regime.



\vfill

\end{titlepage}

\section*{Introduction}

There is growing interest in the dynamics of higher spin states in String Theory \cite{ZNS2, Bianchi:2006nf, Ko:2008ft, Feng:2010yx, Bianchi:2010es, Schlotterer:2010kk}. Although only under very special conditions states of these kind can saturate a BPS bound and thus be perturbatively stable \cite {Bianchi:2010es, Bianchi:2010dy, TesiLucaL}, higher spin states are a hallmark of String Theory or a by-product of strongly coupled gauge theories, possibly captured by some version of the holographic correspondence \cite{Maldacena:1997re, Gubser:1998bc, Witten:1998qj, Aharony:1999ti}.

We here consider the scattering of higher spin closed-string states off D-branes \cite{Lee:2011ita}. For states in the `first Regge trajectory' we find remarkable simplifications both at tree level and at one loop at high energy $\ap E^2>>1$. This parallels the recent analysis for massless states in the Regge limit \cite{D'Appollonio:2010ae}, where the validity of the eikonal approximation was checked by comparing tree-level and one-loop amplitudes in the high energy limit. A semi-classical picture emerges that suggests a resummation is possible of the dominant terms in the amplitudes, conjectured in \cite{D'Appollonio:2010ae} to scale as $E^{n+1}$ for surfaces with $n$ boundaries. Subdominant tidal effects due to the finite length of the string have been also studied in \cite{D'Appollonio:2010ae}.

As in \cite{D'Appollonio:2010ae}, we will eventually focus on the Regge limit of large $s$ and small scattering angle $|t|<< s$. We will only consider massive higher spin states with $M^2 << s$ \ie at fixed level, starting in fact with the first non-trivial case $N=\bar{N}=2$. As we will see later on, amplitudes for the production of a massive higher spin state from a massless graviton impinging on a stack of Dp-branes grow as $E^2$ at tree level and $E^3$ at one loop, independently of the spin of the massive product. One effectively regains the same high energy behavior as in the massless case for $E^2 >> M^2, |t|$, thus corroborating the eikonal description advocated by \cite{D'Appollonio:2010ae}.

It is important to notice that while in the Regge limit (small deflection angles) amplitudes diverge with $E$, at {\it fixed} scattering angle string amplitudes are safely exponentially suppressed in the UV \cite{Green:1987mn, Green:1987sp, Gross:1987kza, Gross:1987ar}. There is a vast literature on trans-Planckian collisions of gravitons in String Theory \cite{Amati:1987wq, Amati:1987uf, Amati:1988tn, Amati:1990xe, Amati:1993tb, Amati:1992zb} that led to important insights in the understanding of the modification of space-time geometry at short distances.

The plan of the paper is as follows.

In Section \ref{VertOps}, we present vertex operators for closed-string higher spin states both in the bosonic string and in the superstring. In particular we identify BRST invariant vertex operators for states in the first Regge trajectory.

In Section \ref{DiskAmp}, after reviewing some old and recent results, we compute tree-level amplitudes on the disk involving both massless states and massive higher spin states in the first Regge trajectory. We then extrapolate the result to very high energy in the Regge limit.

In Section \ref{AnnulAmp} we compute one-loop amplitudes on the annulus involving massive higher spin states. For states in the first Regge trajectory, we find remarkable simplifications in the high energy limit that allow explicit summation over spin structures in a very similar way to what happens for massless states. Moreover the result can be recast in terms of a sum of convolutions of disk amplitudes in these regime.

In Section \ref{ReggeLimEtc} we comment on the compatibility of our findings with the eikonal operator proposed in \cite{D'Appollonio:2010ae}.

Section \ref{Conclus} contains a summary of the present results and our conclusions.

We have collected some cumbersome formulae in an Appendix.

{\it Note added}

While this paper was being typed, a very interesting paper appeared \cite{Black:2011ep} that  overlaps with our analysis at tree level and presents an interpretation in terms of OPE and Pomeron vertex but does not discuss one-loop corrections\footnote{See also \cite{Fotopoulos:2010cm} for a recent discussion of the role of `Pomerons' in deriving recursion relations for scattering amplitudes on D-branes.}.

\section{Vertex operators for Higher Spins}
\label{VertOps}

In this section we will present BRST invariant vertex operators for massive higher spin states of the Bosonic String as well as of the Superstring. We will mainly focus on states belonging to the leading Regge trajectory of the graviton. For notational simplicity we will set $\ap = 2$ henceforth, unless explicitly indicated. We will also try to consistently denote massless momenta by $k$'s and massive ones by $p$'s.

\subsection{Bosonic String}
\label{BoseVertOps}

The un-integrated massless vertex operator for closed bosonic strings reads \cite{Green:1987mn, Green:1987sp, Friedan:1985ge}
\be
V_{m=0} = M_{\m\n}  \de
 X^\mu \bar\de X^\nu e^{ik\cdot X}
\label{m0bosvert}\ee
BRST invariance requires $k^2 = 0$ and $k^\mu M_{\mu\nu}=0$. The polarization tensor  $M_{\mu\nu}$ describes gravitons ($M_{\mu\nu} = M_{\nu\mu}$ and $\eta^{\mu\nu} M_{\nu\mu} = 0$), anti-symmetric tensors ($M_{\mu\nu} = - M_{\nu\mu}$) aka Kalb-Ramond fields, and dilatons ($M_{\mu\nu} = \eta_{\mu\nu} - k_\mu \hat{k}_\nu - k_\nu \hat{k}_\mu$ with $\hat{k}^2 = 0$ and $k\hat{k}= 1$).

Massive vertex operators are given by \cite{Bianchi:2006nf, Bianchi:2010es, TesiLucaL}
\be
V_{M\neq 0} = H_{\m_1\mu_2...\mu_N\nu_1\nu_2...\nu_{\bar{N}}}
\de^{\ell_1} X^{\mu_1} \de^{\ell_2} X^{\mu_2} ... \de^{\ell_N} X^{\mu_N} \bar\de^{\bar\ell_1} X^{\nu_1} \bar\de^{\bar\ell_2} X^{\nu_2} ... \bar\de^{\bar\ell_{\bar{N}}} X^{\nu_{\bar{N}}} e^{ipX} \ee
with $- p^2 = M^2 = 2(N-1) = 2(\bar{N}-1)$ for level matching and $N = \Sigma_i \ell_i$, $ \bar{N}= \Sigma_j \bar\ell_j$. In general, conditions for BRST invariance are very involved and tend to relate different vertex operators of the above kind. However for states in the `first Regge trajectory'  $\ell_i = 1 = \bar\ell_j$ for all $i$ and $j$ the situation drastically simplifies. In particular, at level $N=\bar{N}$, the highest spin state with $J=2N$ is described by a totally symmetric tensor $H_{(\mu_1\mu_2...\mu_N \nu_1\nu_2...\nu_N)}$, subject to the BRST conditions
\be
p^{\mu_1} H_{(\mu_1\mu_2...\mu_N \nu_1\nu_2...\nu_N)} = 0 \quad , \quad  \eta^{\mu_1\mu_2} H_{(\mu_1\mu_2...\mu_N \nu_1\nu_2...\nu_N)} = 0
\ee

Since the number of states grows exponentially with $\sqrt{N}$, it is very hard if not even hopeless to find a systematic way to describe BRST invariant states in a covariant fashion\footnote{See however \cite{Skliros:2011si} for a recent thorough analysis with applications to cosmic strings}. It is straightforward to identify all the physical states in the light-cone gauge, that only exposes $SO(24)$ symmetry, the little Lorentz group for massless momenta in $D=26$. Level by level, it is relatively easy to assemble massive states into representations of $SO(25)$, the little Lorentz group for massive momenta in $D=26$ \cite{Hanany:2010da, Beisert:2004di, Bianchi:2003wx, Bianchi:2004xi}.

At the first massive level $N=\bar{N}=2$, for instance, one finds $24 + (1/2)\cdot 24 \cdot 25$ states for the left-movers, that correspond to the $(1/2)\cdot 26 \cdot 25 -1 $ physical polarizations of a massive symmetric and traceless tensor in $D=26$. Combining with the right-movers yields not only totally symmetric tensors of spin ranging from $J_{Max}=4$ to $J_{min} =0$ but also tensors with mixed symmetry.

\subsection{Superstring}
\label{SuperVertOps}

We will now focus on bosonic states in the NS-NS sector of the closed Type II A or B superstrings. We will neither discuss bosonic states in the R-R sector nor fermionic states in the R-NS and NS-R sectors. Though interesting by themselves, they should not add anything significantly new to our analysis except for a possible direct check of supersymmetry at higher mass levels.

In the canonical superghost picture $q=\bar{q}=-1$, un-integrated massless vertex operators for bosonic NS-NS states read \cite{Green:1987sp, Green:1987sp, Friedan:1985ge}
\be
V^{(-1,-1)}_{m=0} = M_{\mu\nu} \Psi^\mu \bar\Psi^\nu  e^{-\varphi - \bar\varphi}  e^{ikX}
\ee
Very much as for the bosonic string, BRST invariance requires $k^2 = 0 = k^\mu M_{\mu\nu}$. The polarization tensor  $M_{\mu\nu}$ describes gravitons, anti-symmetric tensors, and dilatons.

In the canonical superghost picture $q=\bar{q}=-1$, un-integrated massive vertex operators  for NS-NS states in the `first Regge trajectory' read \cite{Bianchi:2010es, Bianchi:2010dy, TesiLucaL}
\be
V^{(-1,-1)}_{M\neq 0} = H_{(\mu\mu_1\mu_2...\mu_N \nu \nu_1\nu_2...\nu_{\bar{N}})} \Psi^\mu  \de X^{\mu_1} ... \de X^{\mu_N}
\bar\Psi^\nu \bar\de X^{\nu_1}... \bar\de X^{\nu_{\bar{N}}} e^{-\varphi - \bar\varphi} e^{ipX}
\ee
with $M^2 = - p^2 = 2(N-1) = 2(\bar{N}-1)$. As for the bosonic string, conditions for BRST invariance are very involved and tend to relate different vertex operators of the above kind. For spin $J=2N$ totally symmetric tensors $H_{(\mu_1\mu_2...\mu_N \nu_1\nu_2...\nu_N)}$, the conditions simply read
\be
p^{\mu_1} H_{(\mu_1\mu_2...\mu_N \nu_1\nu_2...\nu_N)} = 0 \quad , \quad  \eta^{\mu_1\mu_2} H_{(\mu_1\mu_2...\mu_N \nu_1\nu_2...\nu_N)} = 0
\ee

Very much as for the bosonic string, the exponential growth with $\sqrt{N}$ of the number of states prevents one from finding a systematic way to describe BRST invariant states in a covariant fashion. It is rather straightforward to identify all the physical states in the light-cone gauge with manifest $SO(8)$ symmetry, the little Lorentz group for massless momenta in $D=10$. Level by level, one can assemble massive states into representations of $SO(9)$, the little Lorentz group for massive momenta in $D=10$ \cite{Hanany:2010da, Beisert:2004di, Bianchi:2003wx, Bianchi:2004xi}.

\section{Tree level disk amplitudes}
\label{DiskAmp}

After reviewing tree-level amplitudes for scattering of massless states off D-branes, we will discuss amplitudes with massive higher spin states.

Scattering of closed-string states in the NS-NS sector of the superstring off a stack of $N_p$ Dp-branes is coded in disk amplitudes of the form \cite{Garousi:1996ad, Hashimoto:1996bf}\footnote{See also \cite{Becker:2011bw} for a recent analysis of the decoupling of BRST trivial states.}
\be
\cA^{disk} =  i \kappa \cT_p N_p \int {d^2 z_1 d^2 z_2 \over \cV_{CKV}} \langle V_1^{(-1,-1)} V_2^{(0,0)} \rangle
\ee
where $\cV_{CKV}$ is the (infinite) volume of the conformal Killing group of the disk, $\kappa = 2^6 \pi^7 (\ap)^4 g_s^2 = \sqrt{8\pi G_N}$, with $G_N$  the Newton's constant in $D=10$, $g_s$ is the string coupling constant and ${\cT}_p = 1/(2\pi)^p g_s (\ap)^{p+1/2}$ is the Dp-brane tension. Setting $z_1 = i y$ and $z_2 =i$ yields
\be
{d^2 z_1 d^2 z_2 \over \cV_{CKV}} \rightarrow 4(1-y^2) dy
\ee

Wick contractions are performed by means of
\be
\langle X^\mu(z) X^\nu(w) \rangle = - \eta^{\mu\nu} \log(z-w)
\qquad , \qquad \langle \Psi^\mu(z) \Psi^\nu(w) \rangle = {\eta^{\mu\nu}\over z-w}
\ee
Eventually the boundary condition will convert R-movers $\tilde\Phi^\mu(\bar{z})$ into L-movers $\bar\Phi^\mu(\bar{z}) = R^\mu{}_\nu\Phi^\nu(\bar{z})$ with
\be
R^\mu{}_\nu = {\rm diag}(+1,..+1,-1,...-1)
\ee
the Dp-brane reflection matrix with signature $(p+1,9-p)$. Notice that $RR=1$ and $R^t=R$. Moreover we will frequently use the compact notation $\bar{v}^\mu = R^\mu{}_\nu v^\nu$.

Since we will only consider totally symmetric transverse tensors it proves convenient in the intermediate steps, \ie while performing Wick contractions, to set
\be
h_{\mu\nu} = A_\mu A_\nu \quad \ie \quad h = A\otimes A
\ee
with $A_\mu$ a (complex) massless vector polarization, with $k\cdot A=0$ and $A\cdot A=0$, and
\be
H_{\mu_1...\mu_N \nu_1...\nu_N} = B_{\mu_1}... B_{\mu_N} B_{\nu_1}...B_{\nu_N} \quad \ie \quad H = B\otimes ... \otimes B
\ee
with $B_\mu$ a (complex) massive vector polarization with $p\cdot B=0$ and $B\cdot B=0$.

Final expressions for tree-level amplitudes boil down to integrals of the form
\be
I(a,b,c)= \int_0^1 dy y^a (1-y)^b (1+y)^c
\ee
If $2a+b+c = -2 $, $I(a,b,c)$ can be mapped into a standard Beta function via
\be
y = {1-\sqrt{x} \over 1+\sqrt{x} } \ee
so that
\be
I(a,b,-2-b-2a) = \int_0^1 dx x^{b-1\over 2} (1-x)^a = 2^{-2-2a} {\cB}(a+1, b+1/2)
\ee
Including the ghost measure factor $(1-y^2)$ and setting $a=-2s - n_3$, $b=4s+t-M^2 - n_1 + 1$ and $c=-t +M^2 - n_2 + 1$, the condition on $n_1$, $n_2$, $n_3$ becomes $n_1 + n_2 + 2n_3 = 4$. At high energy, for a given tensorial structure (kinematical factor) with a fixed power of $s = E^2$, the leading dynamical terms turn out to be the ones with the highest value of $n_1$.

\subsection{Massless to massless}
\label{DiskAmpm0m0}

For massless to massless scattering at tree-level (disk) one finds \cite{Garousi:1996ad, Hashimoto:1996bf}
\be
\cA^{disk}_{00} = -i \kappa {\cT}_p N_p 2^{k_1\bar{k}_1 + k_2\bar{k}_2 + 1} \int_0^1 dy y^{k_2\bar{k}_2} (1-y)^{2k_1k_2} (1+y)^{2k_1\bar{k}_2}
\left[ {K_1 \over 1-y^2} - {K_2 (1-y) \over 4 y(1+y)} \right]
\ee
where $\bar{k}_i^\mu = R^\mu{}_\nu k_i^\nu$ and
\bea
&& K_{1} = - k_{1}\cdot  {h}_{2} \cdot {h}_{1} \cdot  k_{2}
-k_{1} \cdot {h}_{2}^{t} \cdot {h}_{1} \cdot \bar{k}_{1} - k_{1} \cdot {h}_{2} \cdot {h}_{1}^{t} \cdot \bar{k}_{1}
- k_{1} \cdot {h}_{2} \cdot {h}_{1}^{t}  \cdot k_{1} +\nn\\ &&\qquad  + Tr({h}_{1} \cdot {R}) k_{1}\cdot {h}_{2} \cdot k_{1} - s Tr({h}_{1} \cdot {h}_{2}^{t}) + (1 \leftrightarrow 2)
\eea
\bea
&&K_{2} = Tr({h}_{1} \cdot {R})(k_1 \cdot {h}_2 \bar{k}_2 + \bar{k}_2 \cdot {h}_2 \cdot k_1 +
\bar{k}_2 \cdot {h}_2 \cdot \bar{k}_2) +\nn \\ &&\qquad + \bar{k}_1 \cdot {h}_1 \cdot {R} \cdot  {h}_2  \cdot \bar{k}_2
- \bar{k}_2 \cdot {h}_2 \cdot {h}_{1}^{t} \bar{k}_1 +\nn \\  &&\qquad - s Tr({h}_1 \cdot {R} \cdot {h}_2 \cdot {R})
+ s Tr({h}_1 \cdot {h}_{2}^{t}) - Tr({h}_1 \cdot {R})Tr({h}_2 \cdot {R}) (- s - {t \over 4})\nn\\  &&\qquad + (1 \leftrightarrow 2).
\eea
with
\be
t = - (k_1 + k_2)^2 = - 2k_1k_2 \quad , \quad 4 s = - (k_1 + \bar{k}_1)^2 = - 2k_1\bar{k}_1
\ee
Exploiting $SO(p,1) \times SO(9-p)$ symmetry, one can set $k_{1\parallel} = \bar{k}_{1\parallel} = (E,{\bf 0}) = - k_{2\parallel} = - \bar{k}_{2\parallel} $, $k_{1\perp} = ({\bf k}_1) = -\bar{k}_{1\perp}$, $k_{2\perp} = ({\bf k}_2) = - \bar{k}_{2\perp}$ with ${\bf k}_1^2 = E^2 = {\bf k}_2^2$.
Defining ${\bf q}={\bf k}_1+{\bf k}_2$ one then finds
\be
t = - {\bf q}^2 =  - 4 E^2 \sin^2(\vartheta/2)  \quad , \quad s = E^2
\ee
Performing the integrals one finally gets
\be
\cA^{disk}_{00} = i {\kappa {\cT}_p N_p \over (4s + {t})} \left(- 2 s K_1 + K_2 {t\over 2}\right) {\cB}(- 2s, -{t\over 2})
\ee

The Gamma functions in the amplitude produce two infinite series of poles corresponding to closed-string states with $\ap M^2 = 4(N-1)$ in
the $t$-channel and to open-string states with $\ap m^2 = n-1$ in the
$s$-channel.

In the special case where the only non-zero components of ${h}_{i\mu\nu}$ are along the directions transverse to the Dp-brane, the amplitude simplifies drastically. For two gravitons one gets
\be
\cA^{disk}_{00} =  i \kappa \cT_{p} N_p s {\Gamma(- {t \over 2})\Gamma(1 - 2s) \over \Gamma(1- {t \over 2} - 2s)} Tr({h}_1 \cdot {h}_2)
\ee

\subsubsection{Regge Limit for massless to massless scattering}

At high energy, the kinematical factor for two transversely polarized gravitons gives
\be
K^{disk}_{00}(k_1, {h}_1 ; k_2, {h}_2) = - 2 s K_1 + K_2 {t\over 2} \approx Tr({h}_{1}{h}_{2})(\ap s)^{2}
\ee
while, reinstating $\ap$ and normalization constants and exploiting Stirling formula for the asymptotic behavior of the Gamma functions, the dynamical factor yields
\be
B^{disk}_{00} \approx \Gamma(-{\ap\over 4}t)e^{-i\pi{\ap t \over4}}(\ap s)^{- 1 + {\ap t \over 4}}
\ee
so that \cite{D'Appollonio:2010ae}
\be
\cA^{disk}_{00} \approx \kappa \cT_p N_p Tr({h}_{1}{h}_{2}) \Gamma(-{\ap\over 4}t)e^{-i\pi{\ap t \over4}}(\ap s)^{1 + {\ap t \over 4}}
\ee
This shows that the disk amplitude grows as $E^2$ in the Regge limit and is dominated by the Regge trajectory of the graviton in the $t$-channel. In the strict Field Theory limit, $\ap \rightarrow 0$, the disk amplitude
\be
\mathcal{A}^{disk}_{00}\rightarrow \mathcal{A}^{tree}_{FT} =  4 \kappa \cT_p N_p Tr({h}_{1}{h}_{2}) {s \over (-t)}
\ee
describes single massless graviton exchange with transverse momentum transfer $-t = {\bf q}^2$.

The imaginary part of $\cA^{disk}_{00}$ accounts for inelastic processes in which bulk closed-string states excite open-string states on the Dp-brane.
As energy grows, the one-particle exchange approximation violates the unitary bound and one needs to consider multi-particle exchanges that correspond to amplitudes on higher-genus surfaces with many boundaries.

\subsection{Massless to Massive}
\label{DiskAmpm0M}

Let us now consider inelastic processes in which a massless graviton produces a massive higher spin state upon scattering on a stack of Dp-brane. The relevant tree-level amplitude reads
\be
\cA_{0M}^{disk} = i\kappa \cT_p  \int {d^2 z_1 d^2 z_2 \over \cV_{CKV}}  \langle V_{1, m=0}^{(0 , 0)}(z_1 , \bar{z}_1 )V_{2, M\neq 0}^{(-1 , -1)}(z_2 , \bar{z}_2 )\rangle
\ee
where
\be
V_{1, m=0}^{(0 , 0)}  = {h}_{\mu\nu}(\partial X^{\mu}+ ik\cdot\Psi \Psi^{\mu})(\bar{\partial} X^{\nu} + i {k}\cdot\tilde\Psi \tilde\Psi^{\nu})e^{i {k}\cdot X}
\ee
is the massless vertex in the $q=\bar{q}=0$ super-ghost picture, while, for the time being, we take
\be
V^{(-1 , -1)}_{2, M\neq 0 }  = H_{\mu\nu\rho\sigma} e^{\phi}\partial X^{\mu}\Psi^{\nu} e^{-\tilde\phi}\bar{\partial}X^{\rho} \tilde\Psi^{\sigma}  e^{i {p}\cdot X}
\ee
for the massive one. It describes a spin 4 particle state in the first Regge trajectory at the first massive level $N=\bar{N}=2$. Later on we will generalize our results to arbitrary level and spin in the `first Regge trajectory'.
Setting
\be
4s = -(k + \bar{k})^2 = -(p + \bar{p})^2\qquad , \qquad t = -(k  + p)^2 = -(\bar{k} + \bar{p})^2
\ee
and recalling that $p^2 = \bar{p}^2 = - M^2$ and $k^2 = \bar{k}^2 = 0$  yield the following kinematical invariants
\bea
&& k\cdot\bar{k} = -2s = p\cdot\bar{p} - M^2 \nn \\
&& p\cdot k  =  {M^2\over 2}- {t\over 2} = \bar{p}\cdot\bar{k} \nn \\
&& p\cdot\bar{k} = 2s + {t\over 2} - {M^2\over 2} =  \bar{p}\cdot k
\eea
The amplitude can be decomposed into four sub-amplitudes
\be
\mathcal{A}^{disk}_{0,M} = \mathcal{A}^{\psi\tilde\psi, disk}_{0,M} + \mathcal{A}^{\psi^3\tilde\psi, disk}_{0,M} + \mathcal{A}^{\psi\tilde\psi^3, disk}_{0,M} + \mathcal{A}^{\psi^3\tilde\psi^3, disk}_{0,M}
\ee
according to the number of world-sheet fermions involved, \ie two, four, four and six.

Using polarization tensors in factorized form $h_{\mu\nu} = A_\mu A_\nu$, for the graviton, and $H_{\mu\nu\rho\sigma} = B_\mu B_\nu B_\rho B_\sigma$ for the massive spin 4 particle,  performing the tedious Wick contractions and setting $z_1 = i y$ and $z_2 = i$, one arrives at a plethora of integrals of the form
\be
I(n_1,n_2,n_3) = K(n_1,n_2,n_3)\int_{0}^{1} dy (1- y^2 ) {y^{k\cdot \bar{k}} (1-y)^{2p\cdot k} (1+ y)^{2p\cdot\bar{k}}\over y^{n_3} (1-y)^{n_1} (1+y)^{n_2}}
\ee
where $K(n_1,n_2,n_3)$ denote independent kinematic structures \ie contractions of polarizations and momenta. One can easily check that the condition $2n_3 + n_1 + n_2 =  4$, or equivalently $2a + b +c = -2$, for the integrals to be expressible in terms of Beta functions is always satisfied.

The final somewhat cumbersome expressions for the four sub-amplitudes are collected in the  Appendix.

In order to generalize the result to arbitrary level $N>2$ in the leading Regge trajectory, it proves convenient to write (chiral) vertex operators in exponential form
\be
W_N =  e^{ipX} (B\partial X)^{N} = {\partial^N \over \partial \beta^N} \left[ e^{ipX + \beta B\partial X}\right]_{\beta= 0}
\ee
This drastically simplifies the combinatorics \cite{Bianchi:2010es, Bianchi:2010dy, TesiLucaL}. Not surprisingly the final expressions are even more cumbersome than in the $N=2$ case. We refrain from displaying the general formulae which are not very illuminating and instead pass on to consider the Regge limit first for level $N=2$ and then for arbitrary $N$.

\subsubsection{Regge limit at tree level (disk)}
\label{ReggeLimTree}

As already observed, for a `fixed' tensor structure, \ie  for a given power of the energy $E$ in the kinematical factors $K(n_1,n_2,n_3)$, the dominant terms are the ones with largest $n_1$. It turns out that this is (always) achieved when $n_1=J=2N = M^2 + 2$.

For $M^2= 2$ ($N=2$) the dominant term for $\ap s >> 1$ at fixed $\ap t$ comes from the sub-amplitude $\cA^{\psi^3\tilde\psi^3 (disk)}_{0M}$ (see Appendix).  More precisely it obtains from the maximal number of contractions between fermion insertions at $z_1$ ($\bar{z}_1$) and at $z_2$ ($\bar{z}_2$), producing factors of $AB$ and $\bar{A}\bar{B} = AB$, once a contraction between fermions at $z_1$ and at $\bar{z}_1$ has produced a $k\bar{k} = -2s$ factor. Moreover, the leading contribution from the bosonic coordinates is given by contracting $\partial X(z_2)$, respectively $\bar\partial X(\bar{z}_2)$, in the massive vertex operator with $e^{ikX(z_1)}$, respectively with $e^{i\bar{k}X(\bar{z}_1)}$, in the massless vertex operator. These contractions produce factors of $Bk$ and $\bar{B} \bar{k}$. Notice that $\bar{B} \bar{k} = Bk = {\bf B}{\bf q}$ where ${\bf q} = {\bf k}_\perp +  {\bf p}_\perp$ is the momentum transferred to the Dp-brane in the $9-p$ transverse directions, where lack of translation invariance allows for violation of momentum conservation. It is important to observe that $Bk = {\bf B}{\bf q}$ does not grow with $E$ since $|{\bf q}|^2=-t << s=E^2$ in the Regge limit.
The final expression for the leading terms at large $E$ turns out to be
\be
\cA^{(disk)}_{0M} \approx {(k\bar{k})(k\bar{k}-1)(k\bar{k} + kp + 1)\over 16(k p)(kp -1)} (AB)^2 (Bk)^2{\cB}(kp  +1, k\bar{k} - 1)
\ee
Exploiting Stirling formula and replacing $BBBB$ with $H$ and $AA$ with $h$ one eventually finds
\be
\cA_{0,M^2=2}^{disk} \approx \kappa \cT_p N_p (\ap s)^{1+ {\ap t\over 4}} e^{-i\pi{\ap t\over 4}}\Gamma\left(-{1\over 4} \ap t \right) \left({\ap\over2}\right) {h}^{\mu\nu}H_{\mu\nu\rho\sigma}k^{\rho}k^{\sigma}
\ee
which coincides with the corresponding formula for massless to massless scattering up to the replacement of $h_{1\mu\nu} h_{2}^{\mu\nu}$ with $\ap {h}^{\mu\nu}H_{\mu\nu\rho\sigma}k^{\rho}k^{\sigma}$.
Once the rule is clear for the first non-trivial case $N=2$, generalization is almost straightforward to any $N$. At least in the Regge limit, there is no need to perform tedious  contractions. For scattering  of massless states (gravitons) to massive states with $J=2N > 4$, the dominant term comes from $\cA^{\psi^3\tilde\psi^3 (disk)}_{0,M^2=2(N-1)}$ following the same rule as above. It correspond to an integral with $n_1=2N$ that eventually yields
\be
\cA^{(disk)}_{0,M^2=2(N-1)} \approx \kappa \cT_p N_p (\ap s)^{1+ {\ap t\over 4}} e^{-i\pi{\ap t\over 4}}\Gamma\left(-{1\over 4} \ap t \right) \left({\ap\over2}\right)^{N-1} {h}^{\mu_1\mu_2}H_{\mu_1\mu_2\mu_3 ...\mu_{2N}}k^{\mu_3}...k^{\mu_{2N}}
\ee

As expected the dominant high energy behavior is $E^2$, independently of the level $N = J/2 = 1 + (\ap/4) M^2$. This strongly supports the validity of the eikonal approximation in the Regge limit as advocated in \cite{D'Appollonio:2010ae}.

\section{One-loop amplitudes (annulus)}
\label{AnnulAmp}

We will now proceed with the analysis of one-loop amplitudes. For completeness, we will first briefly reviewed the derivation and analysis of the massless to massless amplitude. We will closely follow the discussion in \cite{D'Appollonio:2010ae} in order to isolate the dominant terms in the Regge limit of amplitudes involving massive states in the leading Regge trajectory. Quite remarkably, for such states the relevant fermionic contractions simplify drastically in the high energy limit and one can perform the sum over the spin structures as for the massless case.

\subsection{Massless to massless}

The annulus amplitude admits two descriptions. First as a loop of open strings (direct or loop channel). Second as a boundary-to-boundary tree-level closed string exchange (transverse or tree channel, cylinder). The two are related by an S-modular transformation
$\tau_{op}\rightarrow \tau_{cl} = - {1/\tau_{op}}$. The modular parameter of a torus doubly-covering an annulus is purely imaginary $\tau_{op} = i\lambda $ \cite{Green:1987mn,Green:1987sp}.
Since at large energy and large impact parameter the integral is dominated by the large $\lambda =  Im(\tau_{cl})$ region, it proves convenient to adopt the transverse channel vantage point.
The boundary-to-boundary amplitude with insertion of massless emission vertices reads \cite{D'Appollonio:2010ae}
\be
A_{2}(p_1, {h}_1 ; p_2, {h}_2) = \mathcal{C}_p \int d^2 z_1 d^2 z_2 \langle Dp|V_1 (z_1 , \bar{z}_1)V_2 (z_2 , \bar{z}_2)\Delta |Dp\rangle
\ee
where $\mathcal{C}_p$ is a normalization constant and  $\Delta$  is the closed-string propagator.

Wick contractions are performed by means of the bosonic propagator (Bargmann kernel)
\be
G(z,w) = -{1\over 2}\left[ log{|\theta_{1}(z-w)|^2\over |\theta_{1}^{\prime}(0)|^2} - {2\pi\over Im(\tau )}Im(z -w)^2\right]
\ee
and the fermionic propagator (Szego kernel)
\be
S_{\alpha}(z - w) = {\theta_{\alpha}(z-w)\over\theta_{1}(z-w)}{\theta_{1}^\prime(0)\over\theta_{\alpha}(0)}
\ee
for even spin structure $\alpha$.  Thanks to Riemann identity for Jacobi theta functions,
only contractions of the four fermion bilinears survive summation over the even spin structures.
Setting $z_i = \omega_i + i \rho_i \lambda$ for $i=1,2$ with $0<\omega_i<1$ and $0<\rho_i<1/2$, the massless to massless closed-string amplitude on the annulus can be written in the form \cite{D'Appollonio:2010ae}
\be
\mathcal{A}^{ann}_{00}(s,t) = {2\pi \kappa^2 \cT_p^2 N_p^2 K_{tree} \over (2\pi^2\ap)^{7-p\over 2}}\int_{0}^{\infty}{d\lambda\over
 \lambda^{5-p\over 2}}\int_{0}^{1\over 2}d\rho_1 \int_{0}^{1\over 2}d\rho_2 \int_{0}^{1}d\omega_1 \int_{0}^{1}d\omega_2 \, \mathcal{I}(s,t; \lambda, \rho_i, \omega_i)
\ee
where the kinematical factor $K_{tree} \approx (\alpha^\prime s)^2 (h_1 h_2) + ...$ is the same as at tree level, while the integrand $\cI$ is given by
\be
\mathcal{I} = e^{-\ap s \cV_{s} - {\ap t\over 4}\cV_{t}}
\ee
with
\be
\cV_s = -2\pi\lambda\rho^2 +
\log{\theta_1 (i\lambda(\zeta + \rho))\theta_1 (i\lambda(\zeta - \rho))\over
 \theta_1 (i\lambda\zeta + \omega)\theta_1 (i\lambda\zeta - \omega)}
\ee
and
\be
\cV_t = 8\pi\lambda\rho_1 \rho_2 + \log{ \theta_1 (i\lambda\rho + \omega )\theta_1 (i\lambda\rho - \omega)\over
  \theta_1 (i\lambda\zeta + \omega)\theta_1 (i\lambda\zeta - \omega)}
\ee
where
\bea
&&  z_1 - z_2 = \omega + i\lambda\rho\quad , \quad z_1 - \bar{z}_1 = i\lambda (\rho + \zeta )\quad , \quad z_1 -\bar{z}_2 = \omega + i\lambda\zeta\nn \\
 && \bar{z}_1 - z_2 = \omega - i\lambda\zeta\quad , \quad \bar{z}_1 - \bar{z}_2 = \omega - i\lambda\rho\quad , \quad z_2 - \bar{z}_2 = i\lambda (\zeta - \rho )
 \eea
with $\rho\,=\, \rho_{1}-\rho_{2}$, $\zeta\,=\, \rho_{1}+\rho_{2}$ e $\omega\, = \, \omega_{1}-\omega_{2}$.
Thanks to translation invariance along the boundary `circle' direction, the amplitude is independent from $ \omega_{1}+\omega_{2}$.

\subsubsection{Field Theory limit at one-loop}

In order to study the field theory limit, $\ap \rightarrow 0$, it is convenient to set $T = \ap \lambda$, that is kept `fixed' in this regime, dominated by large values of $\lambda$. In terms of the new integration variables the amplitude reads\cite{D'Appollonio:2010ae}
 \be
\mathcal{A}^{ann}_{00} =  {\kappa^2 \cT_p^2 N_p^2 s^{2}\over \pi^{6-p} 2^{7-p\over 2}} \int_{0}^{\infty}
{dT \over T^{5-p \over 2}}\int_{0}^{1}d\zeta \int_{\cR_\rho(\zeta)} d\rho \int_{0}^{1}d\omega \,  e^{-\ap s \cV_{s} - {\ap t \over 4}\cV_{t}}
\ee
where all explicit powers of $\ap$ have disappeared except for the factor in the exponent and $\cR_\rho(\zeta) =\{ 0<\rho<\zeta: \zeta<1/2, 1-\zeta>\rho>0: \zeta>1/2\}$ is  $\rho$ integration region\footnote{This is slightly at variant wrt \cite{D'Appollonio:2010ae} but will only play a role when considering massive states.}.
For large $\lambda$ one has
\be
\ap \cV_{s} \approx -2\pi T \rho^2\qquad , \qquad  \ap \cV_{t} \approx -2\pi T \zeta (1-\zeta)
\ee
The first estimate suggests that the $\rho$ integral is dominated by the region $\rho \approx 0$ at high energy. In the same limit the integrand turns out to be independent of $\omega$, that can be integrated over trivially. In order to render the $\rho$ integral convergent one has to analytically continue $s=E^2= -k\bar{k}$ to negative values. Resorting to the saddle-point approximation for the integral over $\rho$ and then performing exactly the integrals over $T$ and $\zeta$ the field theory limit of the one-loop massless amplitude is given by
\be
\mathcal{A}_{00}^{(ann)} \approx {i\sqrt{\pi} \over \sqrt{2}} {E^{3}\kappa^2 \cT_p^2 N_p^2
\over \pi^{9-p \over 2} 2^{7-p\over 2}}
\left( {2 \over |t|}\right)^{p-4 \over 2}{\Gamma\left({p-4\over 2}\right)\Gamma^{2}\left({6-p\over 2}\right)\over \Gamma\left(
6-p\right)}
\ee
that grows as $E^3$ at large $E$. Recall that the dominant tree-level term had only an $E^2$ growth. A finer analysis allows to determine the Regge limit of the one-loop amplitude including $\ap$ corrections. We will momentarily describe the procedure for the massless to massive one-loop amplitude highlighting the differences with the massless to massless case considered so far.

\subsection{Massless to massive at one-loop}

The massless to massive annulus amplitude reads
\be
A^{(ann)}_{0,M} = \cC_p \int d^2 z_1 d^2 z_2 \langle Dp| V_{m=0}^{(0,0)}(z_1, \bar{z}_1 )V_{M\ne 0}^{(0,0)}(z_2, \bar{z}_2 )\Delta |Dp \rangle
\ee
where $V_{m=0}^{(0,0)}(z_1 ,\bar{z}_1 )$ is the graviton vertex in the $(0,0)$ picture, while in the same picture the massive one reads \cite{Bianchi:2010es, Bianchi:2010dy, TesiLucaL}
\bea
V_{M\ne 0}^{(0,0)} &=& H_{\mu_{1}....\mu_{N}\nu_{1}....\nu_{N}} \\ &&\times [ \prod_{i=2}^{N}\partial X^{\mu_i } + i p\cdot\Psi \Psi^{\mu_1} \prod_{i=2}^{N}\partial X^{\mu_i } + (N-1) \Psi^{\mu_1} \partial \Psi^{\mu_2}\prod_{i=3}^{N}\partial X^{\mu_i }] \nn \\
&&\times [ \prod_{i=2}^{N}\bar\partial X^{\nu_i } + i p\cdot\tilde\Psi \tilde\Psi^{\nu_1} \prod_{i=2}^{N}\bar\partial X^{\nu_i } + (N-1) \tilde\Psi^{\nu_1} \bar\partial\tilde\Psi^{\nu_2}\prod_{i=3}^{N}\bar\partial X^{\nu_i }]
\nn \eea
Thanks to Riemann identity only terms with fermion bilinears contribute.
Based on our detailed analysis at tree level, we expect the dominant terms in the high energy limit to result from the fermion bilinears with explicit momentum $p$ dependence.
As always we find it convenient to decompose transverse tensor polarizations in terms of transverse  vector ones in the intermediate steps \ie $h=AA$ and $H=B...B$.
Keeping only the dominant terms and performing the Wick contractions yields
\be
\cA^{ann}_{0,M} = {2\pi \kappa^2 \cT_p^2 N_p^2 \hat{K}_{tree} \over (2\pi^2\ap)^{7-p\over 2}} \int_{0}^{\infty}{d\lambda\over
 \lambda^{5-p\over 2}}\int_{0}^{1\over 2}d\rho_1 \int_{0}^{1\over 2}d\rho_2 \int_{0}^{1}d\omega_1 \int_{0}^{1}d\omega_2 \mathcal{J}(\lambda, \rho_i, \omega_i)
\ee
where $\hat{K}_{tree}$ is `formally' the same kinematic factor as for massless states with $A_1=A$ (massless vector polarization) and $A_2=B$ (massive vector polarization). At fixed $k\cdot p$ and $k\cdot\bar{k}$ the integrand $\cJ $ is related to  $\cI$, appearing in the purely massless case, by
\be
\cJ(\lambda, z_i; k_i) = \cI(\lambda, z_i; k_i) \chi^{M^2}(z_2,\bar{z}_2) \langle\langle (B\partial X)^{N-1}(z_2) (\bar{B}\bar\partial X)^{N-1}(\bar{z}_2)\rangle\rangle
\ee
where
\be
\chi(z, \bar{z}) = e^{- \pi {Im(z)^2\over Im\tau}} {\theta_1(z-\bar{z}) \over i \theta_1^\prime(0) }
\ee
while we defined
\be
\langle\langle (B\partial X)^{N-1}(z_2) (\bar{B}\bar\partial X)^{N-1}(\bar{z}_2)\rangle\rangle = {\langle (B\partial X)^{N-1}(z_2) (\bar{B}\bar\partial X)^{N-1}(\bar{z}_2) \prod_i e^{ik_iX(z_i)} \rangle \over
\langle \prod_i e^{ik_iX(z_i)} \rangle}
\ee
whose dominant term in the high energy limit is given by contracting each $\partial X(z_2)$, respectively $\bar\partial X(\bar{z}_2)$, with $e^{ikX(z_1)}$ and $e^{i\bar{k}X(\bar{z}_1)}$, respectively with $e^{i\bar{k}X(\bar{z}_1)}$ and $e^{ikX(z_1)}$, so that
\be
\langle\langle (B\partial X)^{N-1}(z_2) (\bar{B}\bar\partial X)^{N-1}(\bar{z}_2) \rangle\rangle \approx (Bk)^{N-1} (\bar{B}\bar{k})^{N-1}|\partial G(z_2-z_1) - \partial G(z_2-\bar{z}_2)|^{2(N-1)}
\ee

\subsection{Regge limit at one-loop (massless-to-Massive) }
\label{ReggeLimLoop}

At high energies, after analytic continuation to negative $s=E^2$, the $\rho$  integral is dominated by the region $\rho\approx 0$, while the dominant contribution in the integral over $\lambda$ comes from the region of large $\lambda$.

In order to compare with the estimate obtained by \cite{D'Appollonio:2010ae} for the purely massless case, one has to study the correction terms deriving from bosonic contractions \ie powers of derivatives of Bargmann kernels and its derivatives.

For the first correction term $\langle\langle (B\partial X)^{N-1} (\bar{B}\bar\partial X)^{N-1}\rangle\rangle$, one needs
\be
\partial_{z}G(z) = \partial_{z}\log\left[\, sen(\pi z) \prod_{n}(1 - 2cos(2\pi z)q^n + q^{2n} )\right] - {2\pi Im(z)\over i Im\tau}
\ee
where $q = e^{-2\pi i\tau} = e^{-2\pi \lambda}$ and $z = z_1 - z_2 = \omega + i\rho\lambda$ or $z = z_2 - \bar{z}_2 = i\lambda (\zeta - \rho )$.
Taking into account that the dominant contributions to the integral in the relevant limit come from the region of small $\rho$ and large $\lambda$, one finds
\be
|\partial G(z_2 - z_1) - \partial G(z_2 - \bar{z}_2)|^{2(N-1)} \approx \pi^{2(N-1)} \sum_{n=0}^{N-1} \left(^{N-1}_{\: n} \right) \cot(\pi \omega)^{2n} (1-2\zeta)^{2(N-1-n)} \ee
Note that it is crucial that $1>\zeta-\rho>0$ for the approximation to be valid. This happens when the integration over $\rho$ is restricted to the region $\cR_\rho(\zeta)$.

For the other correction factor $\chi^{M^2}$, using the product expansions of $\theta_{1}(z)$ and its derivative, one finds
\be
\chi(z_2 - \bar{z}_2) = e^{- \pi\lambda (\zeta - \rho )^2} {\theta_1(i \lambda (\zeta - \rho)) \over i \theta^\prime(0)} \approx {1\over 2\pi} e^{- \pi\lambda [(\zeta - \rho )^2 - (\zeta - \rho )]}
\ee

Before performing the various integrals, one has to reconsider the approximations of $\cV_s$ and  $\cV_t$ used in order to recover the field theory in the purely massless case. Including terms that would vanish for $\ap \rightarrow 0$ one has \cite{D'Appollonio:2010ae}
\be
\cV_s  \approx -2\pi \lambda \rho^2 - 4[\sin^2(\pi\omega) + \sinh^2(\pi\lambda\rho)]
(e^{- 2\pi\lambda\zeta} + e^{- 2\pi\lambda(1-\zeta) }) \ee
and
\be
\cV_t \approx -2\pi \lambda [\zeta(1-\zeta) +\rho^2] + \log[4(\sin^2(\pi\omega) + \sinh^2(\pi\lambda\rho)]
\ee

After analytic continuation to negative $s=E^2 = -{k\bar{k}\over 2}$, the integral over $\rho$ can be performed by saddle-point methods around $\rho\approx 0$ and yields
\be
\int_{\cR_\rho(\zeta)} d\rho  e^{-2\pi \lambda \rho^2 k\bar{k}} \hat\cJ(\rho, \omega, \lambda, \zeta) \approx {i \over \sqrt{2\lambda |k\bar{k}|}} \hat\cJ(\rho=0, \omega, \lambda, \zeta)
\ee
where $\hat\cJ = \cJ e^{+2\pi \lambda \rho^2 k\bar{k}}$.

Expanding the exponential
\be
e^{-4 k\bar{k} \sin^2(\pi \omega) (e^{- 2\pi\lambda\zeta} + e^{- 2\pi\lambda(1-\zeta)})} =
\sum_{\ell_1,\ell_2}^{0,\infty} {[-4 k\bar{k} \sin^2(\pi \omega)]^{\ell_1+\ell_2} \over \ell_1 ! \ell_2 !} e^{- 2\pi\lambda[\ell_1 \zeta + \ell_2 (1-\zeta)]} \ee
integration over $\omega$ can be performed to get
\be
\int_0^1 d\omega \sin(\pi\omega)^{2 kp - 2n + 2\ell_1 + 2\ell_2}  \cos(\pi\omega)^{2n} = {2\over \pi} \cB(\ell_1 +\ell_2 + N-1-n - {t\over 2} + {1\over 2}, n + {1\over 2})
\ee
where $n=0,... N-1$ and use has been made of $M^2 = 2(N-1)$ and $kp-(N-1) = -t/2$ ($\ap =2$).

Next one can perform the integral over $\lambda$
\be
\int_0^\infty d\lambda \lambda^{p-6 \over 2} e^{\pi \lambda [\zeta(1-\zeta) t - 2\ell_1 \zeta - 2\ell_2 (1-\zeta)]}  = {\Gamma\left( {p-4\over 2}\right) \over  (2\pi)^{p-4\over 2}  [- \zeta(1-\zeta) t + 2\ell_1 \zeta + 2\ell_2 (1-\zeta)]^{p-4\over 2} }
\ee

The final integral over $\zeta$ can be written as a convolution in transverse momentum space along the $8-p$ directions orthogonal to the `large' momentum ${\bf r} = {\bf k}_\perp - {\bf p}_\perp$
\bea
&&\int_0^1 d\zeta {\Gamma\left( {p-4\over 2}\right) (1-2\zeta)^{2(N-1)} (Bk)^{2(N-1)}\over  (2\pi)^{p-4\over 2}  [-\zeta(1-\zeta) t + 2\ell_1 \zeta + 2\ell_2 (1-\zeta)]^{p-4\over 2} } = \nn\\ &&\qquad \sum_{n=0}^{N-1}(-)^n \left(\begin{array}{c}N -1\\n\end{array}\right)\int {d^{8-p} {\bf{v}} \over (2\pi)^{8-p}} {4 (2\pi\ap)^{8-p\over 2} ({\bf B}{\bf v})^{2(N-1-n)} ({\bf B}({\bf q} -{\bf v}))^{2n} \over (4\ell_1 + {\ap} |{\bf{v}}|^2) (4\ell_2 + {\ap} |{\bf{v}}-{\bf q}|^2)}
\eea
with $|{\bf q}|^2 = |{\bf k}_{\perp}+{\bf p}_{\perp}|^2 = - t$ and use has been made of $Bk= {\bf B}{\bf q}$ (transversality of $H$) and $BB=0$ (tracelessness of $H$).
Replacing the infinite sums over $\ell_1$ and $\ell_2$ by contour integrals according to
\be
\sum_{\ell=0}^\infty {f(\ell) \sigma^\ell \over \ell ! (\ell + \tau)} = - \oint_\cC {dz\over 2\pi i} {f(z) \sigma^z e^{-i\pi z} \Gamma(-z) \over \ell ! (z + \tau)}
\ee
with $\cC$ encompassing the positive real axis and deforming the contour around the poles at
$z_i=-\tau_i$ and the first pole of the Beta function one finally gets
\bea
&&\cA_{0M}^{(ann)} \approx i {\pi^{2-p\over 2} (e^{-i\pi} \ap s)^{\ap t/4} \over \sqrt{2\ap s} (2\pi\ap)^{8-p\over 2}} \sum_{n=0}^{N-1}(-)^n \left(\begin{array}{c}N -1\\n\end{array}\right) \int {d^{8-p} {\bf{v}} \over (2\pi)^{8-p}} (AB)(\bar{A}\bar{B}) \\ &&\quad ({\bf B}{\bf v})^{2(N-1-n)} \Gamma(-{\ap t_1\over 4}) ({\bf B}({\bf q} -{\bf v}))^{2n} \Gamma(-{\ap t_2\over 4})\cB({t_1+t_2-t+1\over 2} + n, N-n - {1\over 2}) + ... \nn
\eea
that grows as $E^3$ for large $E$ as in the massless amplitude, while $...$ denotes sub-leading terms and we have defined $t_1= -|{\bf{v}}|^2$ and $t_2= -|{\bf{v}} - {\bf q}|^2$.

Including all constants and using Gamma function duplication formula, one arrives at the final expression for the dominant terms in the massless to massive annulus amplitude
\be
{\cA^{(ann), dom}_{0M}(s,t) \over 2 E} = {i \over 2} \sum_{M'\le M}\int {d^{8-p} {\bf v} \over (2\pi)^{8-p}}
\hat{V}_2(t_1, t_2, t) {\cA^{(disk), dom}_{0M'}(s,t_1) \over 2 E} {\cA^{(disk), dom}_{M'M} (s,t_2) \over 2 E}
\ee
where $\ap (M')^2= 4n \le \ap (M)^2 = 4(N-1) $ and
\be
\hat{V}_2(t_1, t_2, t) = {\Gamma\left[1 + {\ap\over 2} (t_1 + t_2 - t)\right] \over \Gamma\left[n+1 + {\ap\over 4} (t_1 + t_2 - t)\right] \Gamma\left[N-n + {\ap\over 4} (t_1 + t_2 - t)\right] }
\ee
is a `modified' Reggeon vertex that takes into account the emission of massive higher spin particles. A sum over polarizations $H'=(B')^n \otimes (\bar{B}')^n$ of the intermediate state in the first Regge trajectory is understood.
The explicit form of $\cA^{(disk), dom}_{0M'}$ was given in a previous section, while \bea
&&\cA^{(disk), dom}_{M'M} = {1\over 2} \kappa \cT_p N_p (\ap s)^{1+ {\ap t\over 4}} e^{-i{\ap t\over 4}} \sum_{a,b=0}^{min(n, N-1)} \left(\begin{array}{c}N -1\\a\end{array}\right) \left(\begin{array}{c}N -1\\b\end{array}\right)
\left(\begin{array}{c}n\\a\end{array}\right) \left(\begin{array}{c}n\\b\end{array}\right) \times \nn \\ &&
 a! b! (-\ap/2)^{n+N-1-a-b} (B'B)^{a+1} (\bar{B}' \bar{B})^{b+1} (Bq)^{N-1-a} (\bar{B} \bar{q})^{N-1-b} (B'q)^{n-a} (\bar{B}' \bar{q})^{n-b}
\eea
can also be found in \cite{Black:2011ep} in a slightly different notation.
Our results parallel and generalize the analysis of the purely massless case \cite{D'Appollonio:2010ae}, indicating that the scattering of strings off D-branes at one-loop is dominated in the Regge limit by double exchange of states in the Regge trajectory of the graviton, the `first' or leading Regge trajectory. The dominant term in the annulus amplitude factorizes into a two-particle emission vertex and two disk amplitudes.

\section{Regge limit to all orders and Eikonal Interpretation}
\label{ReggeLimEtc}

The divergence of the amplitudes at large $E$ both at tree-level and at one-loop would signal a violation of unitarity in the Regge limit. However the perfect match between the `square' of the dominant term in the disk amplitude and the dominant term in the annulus amplitude signals the possibility of restoring unitarity \cite{D'Appollonio:2010ae}.

Indeed considering S-matrix elements
\be
\cS = 1 + i\cT = 1 + i{\mathcal{A}\over2E}
\ee
and expanding $\cA= \sum_{{n}=1}^{\infty}\mathcal{A}_{{n}}$ as a series over the number of boundaries, one arrives at specific relations between terms that appear at different orders but scale with the same power of $E$ at large $E$.

In order to expose these relations it is convenient to work in impact parameter space, \ie perform a Fourier transform on the $8-p$ space-like directions transverse to the large momentum ${\bf r} = {\bf k} - {\bf p}$ transferred to the D-brane. As a result convolutions in momentum space become simple products in impact parameter space, parameterized by the $8-p$ dimensional vector ${\bf b}$. Using $2i\cT^{dom}_{ann}(E,{\bf b}) = - [\cT^{dom}_{disk}(E,{\bf b})]^2$ strongly suggest the validity of
\be
\cS(E,{\bf b}) = 1 + i\cT_{disk}(E,{\bf b}) + i\cT_{ann}(E,{\bf b}) + . . . . = 1 + i\cT_{disk}(E,{\bf b}) -{1\over 2}(\cT_{disk}(E,{\bf b}))^2 + ....
\ee
In \cite{D'Appollonio:2010ae} next-to-leading terms in $E$  at one-loop were also computed and showed that
$\mathcal{A}_{ann}^{(sub)}$ and $T_{ann}^{(sub)}$ diverge as $E^2$ and $E$ respectively at large $E$, consistent with the above interpretation.

Given our results for massless to massive scattering, it is tempting to conjecture that the dominant term in the one-loop amplitudes involving massive higher spins belonging to the leading Regge trajectory be expressible as convolutions of two dominant tree-level amplitudes. Even more, if at higher orders, $\mathcal{A}_{{n}}$ (on surfaces with ${n}$ boundaries) were dominated by ${n}$ Reggeized graviton exchanges thus producing an $E^{n+1}$ behaviour, as conjectured in \cite{D'Appollonio:2010ae} for massless states, our results at tree-level and one-loop suggest that the same should be true to all orders for any amplitude involving massive higher spins in the `first' Regge trajectory.

\section{Conclusions}
\label{Conclus}

We have studied scattering of massive higher-spin closed supestring states in the first Regge trajectory (of the graviton) off Dp-brane.
Although complete expressions tend to become unwieldy even at tree-level (disk) and for the first massive level $N=\bar{N}=2$, one can easily isolate the dominant contributions at large $E=\sqrt{s}$ for small deflection angle $|t|<< s$. We have shown that, independently of $N$ \ie the spin of the scattered particle, the leading contribution grows as $E^2$ at tree level and as $E^3$ at one-loop. This lends further support to the conjectured form of the eikonal operator proposed in \cite{D'Appollonio:2010ae}.

It would be very interesting to study both dominant terms at higher order (many boundaries) or the subdominant terms in the amplitudes at one-loop with two arbitrary massive higher spins in the first Regge trajectory. One should keep in mind that the remarkable simplifications that occur for states in the leading Regge trajectory considered here are not expected to take place for generic states at arbitrary level. Even solving the BRST conditions and identifying the `physical' states looks like a formidable task at arbitrary level \cite{Bianchi:2010es, Bianchi:2010dy, TesiLucaL}. In the purely massless case, the sub-leading terms at one-loop have been shown to produce a contribution to the deflection angle that explains  corrections in ${\ell_s / b}$ \cite{D'Appollonio:2010ae}.

Other energy regimes could be considered. At fixed deflection angle string amplitudes should decay exponentially with the energy at any order \cite{Green:1987mn, Green:1987sp, Gross:1987kza, Gross:1987ar}. At very small energy one should instead recover the near horizon AdS-like geometry \cite{Gubser:1997qr, Callan:1996tv}. This presents some subtleties in the massive case since it introduces another adimensional parameter, essentially the string level $N$, that played a marginal role -- except possibly for the combinatorics -- in the Regge limit whereby $s >> M^2 > |t|$. Alternatively one could consider the limit $s \approx M^2 >> |t|$. We hope to investigate these and related issues in the near future.

\section*{Acknowledgements}

We would like to thank F.~Fucito, G.~D'Appollonio,P.~Di Vecchia,  L.~Lopez, L.~Martucci, F.~Morales, R.~Richter, R.~Russo and G.~Veneziano for useful discussions. The work of M.~B. was partially supported by the ERC Advanced Grant n.226455 "Superfields", by the Italian MIUR-PRIN contract 20075ATT78,
by the NATO grant PST.CLG.978785. M.~B. would like to thank IMPU, Tokyo and NORDITA, Stockholm for their hospitality during during completion of this project.

\newpage

\section*{Appendix: disk sub-amplitudes}
In this Appendix we collect the four tree-level sub-amplitudes for a massless graviton with $h_{\mu\nu} = A_\mu A_\nu$ and $k^2=0$ to produce a massive spin 4 particle with $H_{\mu\nu\rho\sigma} = B_\mu B_\nu B_\rho B_\sigma$ and $M^2=-p^2=2$ ($\ap =2$). Recall that $\bar{v} = R^\mu{}_\nu v^\nu$ for any vector $v$. A common overall normalization constant is understood. The sub-amplitudes correspond to the number of world-sheet fermions involved in the contractions.

\bea
&& \mathcal{A}^{\psi^3\tilde\psi, disk}_{0,M} = \{{1\over 4}{(k\bar{k} - 1)k\bar{k}\over (k\bar{k} +
kp)(k\bar{k} + kp + 1)}[(A\bar{B})(\bar{A}B)(Bk)(\bar{B}k) + \nn \\
&&  - (AB)(\bar{A}B)(\bar{B}k)(\bar{B}k)] + {1\over 4}{(k\bar{k} -
1)k\bar{k}\over (k\bar{k} + kp)kp}[(A\bar{B})(\bar{A}B)(Bk)(\bar{B}\bar{k})
+ \nn\\
&&  - (AB)(\bar{A}B)(\bar{B}k)(\bar{B}\bar{k})] - {1\over 4}{(k\bar{k} -
1)k\bar{k}\over kp(kp -1)}[(A\bar{B})(\bar{A}\bar{B})(Bk)(Bk) + \nn \\
&& - (AB)(\bar{A}\bar{B})(Bk)(\bar{B}k)] - {1\over 4}{(k\bar{k} -
1)k\bar{k}\over (k\bar{k} + kp)kp}[(A\bar{B})(\bar{A}\bar{B})(Bk)(B\bar{k})]
+ \nn\\
&& - {1\over 16}{k\bar{k} -1\over k\bar{k} +
kp}[(A\bar{B})(B\bar{B})(Bk)(\bar{A}p) - (AB)(B\bar{B})(\bar{A}p)(\bar{B}k)]
+ \nn\\
&& -{1\over 16}{k\bar{k} -1\over
kp}[(A\bar{B})(B\bar{B})(\bar{A}\bar{p})(Bk) -
(AB)(B\bar{B})(\bar{A}\bar{p})(\bar{B}k)]\}\cB(kp + 1 , k\bar{k} - 1)
\eea

\bea
&& \mathcal{A}^{\psi\tilde\psi^3, disk}_{0,M} = \{{1\over 4}{(k\bar{k} - 1)k\bar{k}\over (k\bar{k} +
kp)kp}[(AB)(\bar{A}\bar{B})(B\bar{k})(\bar{B}k) + \nn\\
&& - (AB)(\bar{A}B)(\bar{B}k)(\bar{B}\bar{k}) -
(A\bar{B})(\bar{A}\bar{B})(Bk)(B\bar{k}) +
(AB)(\bar{A}B)(Bk)(\bar{B}\bar{k})] + \nn \\
&&  + {1\over 4}{(k\bar{k} - 1)k\bar{k}\over (kp -
1)kp}[(AB)(\bar{A}\bar{B})(\bar{B}\bar{k})(B\bar{k}) -
(AB)(\bar{A}B)(\bar{B}\bar{k})(\bar{B}\bar{k})] +\nn\\
&& - {1\over 4}{(k\bar{k} - 1)k\bar{k}\over (k\bar{k} + kp)(k\bar{k} + kp
+1)}[(A\bar{B})(\bar{A}\bar{B})(B\bar{k})(B\bar{k}) +\nn\\
&& - (A\bar{B})(\bar{A}B)(B\bar{k})(\bar{B}\bar{k})] + {1\over 16}{k\bar{k}
- 1\over kp}[(\bar{A}\bar{B})(B\bar{B})(Ap)(B\bar{k}) + \nn\\
&& - (\bar{A}B)(B\bar{B})(Ap)(B\bar{k}) -
(\bar{A}B)(B\bar{B})(Ap)(\bar{B}\bar{k})] + {1\over 16}{k\bar{k} - 1\over
k\bar{k} + kp}\cdot \nn\\
&& \cdot [(\bar{A}\bar{B})(B\bar{B})(Ap)(B\bar{k}) -
(\bar{A}B)(B\bar{B})(Ap)(\bar{B}\bar{k})\} \cB(kp + 1 , k\bar{k} - 1)
\eea

\bea
&& \mathcal{A}^{\psi\tilde\psi, disk}_{0,M} = \{{1\over 64}[(Ap)(\bar{A}p)(Bk)(\bar{B}k)(B\bar{B}) +
(Ap)(\bar{A}p)(B\bar{k})(\bar{B}\bar{k})(B\bar{B}) + \nn \\
&& +(Ap)(\bar{A}\bar{p})(B\bar{k})(\bar{B}k)(B\bar{B}) +
(A\bar{p})(\bar{A}p)(Bk)(\bar{B}\bar{k})(B\bar{B}) +
(A\bar{p})(\bar{A}\bar{p})(Bk)(\bar{B}k)(B\bar{B})\nn \\
&& +(A\bar{p})(\bar{A}\bar{p})(B\bar{k})(\bar{B}\bar{k})(B\bar{B})]
 + {1\over 64}{(kp + k\bar{k} -2)(kp + k\bar{k} -1)\over (kp)(kp - 1)}
(Ap)(\bar{A}\bar{p})(Bk)(\bar{B}\bar{k})(\bar{B}B) +\nn \\
&& + {1\over 64}{kp +\bar{k}k -1\over
kp}[(Ap)(\bar{A}p)(Bk)(\bar{B}\bar{k})(\bar{B}B) +
(Ap)(\bar{A}\bar{p})(Bk)(\bar{B}k)(\bar{B}B) + \nn\\
&& + (Ap)(\bar{A}\bar{p})(B\bar{k})(\bar{B}\bar{k})(\bar{B}B) +
(A\bar{p})(\bar{A}\bar{p})(Bk)(\bar{B}\bar{k})(\bar{B}B)] +\nn \\
&& + {1\over 64}{kp +1\over kp +
k\bar{k}}[(Ap)(\bar{A}p)(B\bar{k})(\bar{B}k)(\bar{B}B) +
(A\bar{p})(\bar{A}p)(Bk)(\bar{B}k)(\bar{B}B) +\nn \\
&& + (A\bar{p})(\bar{A}p)(B\bar{k})(\bar{B}\bar{k})(\bar{B}B) +
(A\bar{p})(\bar{A}\bar{p})(B\bar{k})(\bar{B}\bar{k})(\bar{B}B)] +\nn \\
&& + {1\over 64}{(kp + 2)(kp + 1)\over (kp + k\bar{k} +1)(k\bar{k} +
kp)}(A\bar{p})(\bar{A}p)(B\bar{k})(\bar{B}k)(\bar{B}B) +\nn \\
  && + {1\over 64}[(A\bar{A})(Bk)(\bar{B}k)(B\bar{B})
  +(A\bar{A})(B\bar{k})(\bar{B}\bar{k})(B\bar{B})] +\nn \\
&& + {1\over 64}{k\bar{k} + kp -1\over
kp}(A\bar{A})(Bk)(\bar{B}\bar{k})(B\bar{B}) + {1\over 64}{kp +1\over
k\bar{k} + kp}(A\bar{A})(B\bar{k})(\bar{B}k)(B\bar{B}) +\nn \\
&& - {1\over 16}{k\bar{k} - 1\over
kp}[(AB)(\bar{A}p)(\bar{B}\bar{k})(AB)(B\bar{B}) +
(\bar{A}\bar{p})(\bar{B}k)(AB)(B\bar{B})] +\nn\\
&& - {1\over 16}{k\bar{k} -1\over k\bar{k} + kp}
(\bar{A}p)(\bar{B}k)(AB)(B\bar{B})
 - {1\over 16}{(k\bar{k} -1)(k\bar{k} + kp -1)\over kp(kp -1)}
(\bar{A}\bar{p})(\bar{B}\bar{k})(AB)(B\bar{B}) +\nn \\
&& +{1\over 16}{k\bar{k} -1\over kp +
k\bar{k}}[(A\bar{B})(\bar{A}p)(Bk)(B\bar{B}) +
(A\bar{B})(\bar{A}\bar{p})(B\bar{k})(B\bar{B})] +
 {1\over 16}{k\bar{k} -1\over kp}(A\bar{B})(\bar{A}\bar{p})(Bk)(B\bar{B}) +
\nn\\
&& + {1\over 16}{(kp + 1)(k\bar{k} -1)\over ( k\bar{k} + kp)( k\bar{k} + kp
+ 1)}(A\bar{B})(\bar{A}p)(B\bar{k})(B\bar{B})
 + {1\over 16}{k\bar{k} - 1\over kp}(Ap)(\bar{B}\bar{k})(\bar{A}B)(B\bar{B})
+ \nn \\
&& + {1\over 16}{k\bar{k} - 1 \over k\bar{k} +
kp}[(Ap)(\bar{B}k)(\bar{A}B)(B\bar{B}) +
(A\bar{p})(\bar{B}\bar{k})(\bar{A}B)(B\bar{B})] + \nn \\
&& + {1\over 16}{( kp + 1)( k\bar{k} - 1)\over (k\bar{k} + kp )( k\bar{k} +
kp)( k\bar{k} + kp +1)}(A\bar{p})(\bar{B}k)(\bar{A}B)(B\bar{B}) +\nn \\
&& -  {1\over 16}{k\bar{k} - 1\over
kp}[(Ap)(B\bar{k})(\bar{A}\bar{B})(B\bar{B}) +
(A\bar{p})(Bk)(\bar{A}\bar{B})(B\bar{B})] + \nn\\
&& -  {1\over 16}{k\bar{k} - 1\over kp +
k\bar{k}}(A\bar{p})(B\bar{k})(\bar{A}\bar{B})(B\bar{B}) -
  {1\over 16}{( k\bar{k} - 1)(k\bar{k} + kp -1)\over kp( kp
-1)}(Ap)(Bk)(\bar{A}\bar{B})(B\bar{B}) + \nn \\
&& -  {1\over 64}[(Ap)(\bar{A}p)(B\bar{B})(B\bar{B}) +
(A\bar{p})(\bar{A}\bar{p})(B\bar{B})(B\bar{B})] +
 -  {1\over 64}{k\bar{k} + kp - 1\over
kp}(Ap)(\bar{A}\bar{p})(B\bar{B})(B\bar{B}) + \nn \\
&& -  {1\over 64}{ kp + 1\over k\bar{k} +
kp}(A\bar{p})(\bar{A}p)(B\bar{B})(B\bar{B}) + {1\over
64}(A\bar{A})(B\bar{B})(B\bar{B})
 {1\over 4}{ k\bar{k}(k\bar{k} - 1)\over kp(kp -
1)}(A\bar{A})(B\bar{B})(B\bar{B}) + \nn \\
&& +  {1\over 4}{k\bar{k}(k\bar{k} - 1)\over (k\bar{k} + kp)(k\bar{k} + kp +
1)}(A\bar{A})(B\bar{B})(B\bar{B})\} \cB(kp + 1 , k\bar{k} - 1)
\eea

\bea
&& \mathcal{A}^{\psi^3\tilde\psi^3, disk}_{0,M} =  \{ {1\over
64}[(k\bar{k})(Bk)(\bar{B}k)(A\bar{A})(B\bar{B}) -
(Bk)(\bar{B}k)(\bar{A}k)(A\bar{k})(B\bar{B}) + \nn \\
&& (k\bar{k})(B\bar{k})(\bar{B}\bar{k})(A\bar{A})(B\bar{B}) -
(B\bar{k})(\bar{B}\bar{k})(\bar{A}k)(A\bar{k})(B\bar{B}) +
(k\bar{k})(A\bar{A})(B\bar{B})(B\bar{B}) + \nn\\
&& - (A\bar{k})(\bar{A}k)(B\bar{B})(B\bar{B})] + \nn \\
&& -{1\over 16}{k\bar{k} - 1\over
kp}[(AB)(\bar{A}k)(\bar{B}\bar{k})(Bk)(\bar{B}k) +
(Bk)(Bk)(\bar{B}k)(A\bar{k})(\bar{A})(\bar{B}) + \nn \\
&& - (A\bar{A})(Bk)(Bk)(\bar{B}k)(\bar{B}\bar{k}) -
(k\bar{k})(AB)(\bar{A}\bar{B})(\bar{B}k)(Bk) +
(AB)(\bar{A}k)(B\bar{k})(\bar{B}\bar{k})(\bar{B}\bar{k}) + \nn\\
&& + (\bar{A}\bar{B})(A\bar{k})(Bk)(B\bar{k})(\bar{B}\bar{k}) -
(A\bar{A})(Bk)(B\bar{k})(\bar{B}\bar{k})(\bar{B}\bar{k}) -
(k\bar{k})(AB)(\bar{A}\bar{B})(B\bar{k})(\bar{B}\bar{k}) \nn \\
&& + (B\bar{B})(AB)(\bar{A}k)(\bar{B}\bar{k}) +
(B\bar{B})(\bar{A}\bar{B})(Bk)(A\bar{k})  -
(B\bar{B})(A\bar{A})(Bk)(\bar{B}\bar{k}) -
(k\bar{k})(B\bar{B})(\bar{A}\bar{B})(AB)] \nn \\
&&  - {1\over 16}{k\bar{k} - 1 \over k\bar{k} + kp
}[(A\bar{A})(Bk)(B\bar{k})(\bar{B}k)(\bar{B}k) +
(k\bar{k})(A\bar{B})(\bar{A}B)(Bk)(\bar{B}k) +\nn \\
&&  - (A\bar{B})(Bk)(B\bar{k})(\bar{A}k)(\bar{B}k) -
(\bar{A}B)(A\bar{k})(Bk)(\bar{B}k)(\bar{B}k) +
(A\bar{A})(B\bar{k})(B\bar{k})(\bar{B}k)(\bar{B}\bar{k}) + \nn \\
&&  + (k\bar{k})(A\bar{B})(\bar{A}B)(B\bar{k})(\bar{B}\bar{k}) -
(A\bar{B})(B\bar{k})(B\bar{k})(\bar{A}k)(\bar{B}\bar{k}) -
(\bar{A}B)(A\bar{k})(B\bar{k})(\bar{B}\bar{k})(\bar{B}k) +\nn \\
&&  + (A\bar{A})(B\bar{B})(B\bar{k})(\bar{B}k) +
(k\bar{k})(\bar{A}B)(B\bar{B})(A\bar{B}) -
(B\bar{B})(A\bar{B})(\bar{A}k)(B\bar{k}) -
(B\bar{B})(\bar{A}B)(A\bar{k})(\bar{B}k)] + \nn \\
&& +  {1\over 64}{kp +1\over kp +
k\bar{k}}[k\bar{k}(A\bar{A})(B\bar{B})(B\bar{k})(\bar{B}k) -
(A\bar{k})(\bar{A}k)(\bar{B}k)(B\bar{k})(B\bar{k})] + \nn \\
&&  - {1\over 16}{k\bar{k} - 1\over k\bar{k} +
kp}[(AB)(\bar{A}k)(B\bar{k})(\bar{B}k)(\bar{B}\bar{k}) +
(\bar{A}\bar{B})(B\bar{k})(Bk)(A\bar{k})(\bar{B}k) + \nn \\
&&  - (A\bar{A})(Bk)(B\bar{k})(\bar{B}k)(\bar{B}\bar{k}) -
(k\bar{k})(AB)(\bar{A}\bar{B})(B\bar{k})(\bar{B}k)] + \nn \\
&&  - {1\over 16}{(kp +1)(k\bar{k} -1)\over (kp + k\bar{k})(kp + k\bar{k}
+1)}[(A\bar{A})(B\bar{k})(B\bar{k})(\bar{B}k)(\bar{B}k) +
(k\bar{k})(A\bar{B})(\bar{A}B)(B\bar{k})(\bar{B}k)+\nn\\
&& -  (A\bar{B})(\bar{A}k)(B\bar{k})(B\bar{k})(\bar{B}k) -
(\bar{A}B)(A\bar{k})(B\bar{k})(\bar{B}k)(\bar{B}k)] + \nn \\
&&  + {1\over 64}{k\bar{k} + kp - 1\over
kp}[(k\bar{k})(A\bar{A})(B\bar{B})(Bk)(\bar{B}\bar{k}) -
(B\bar{B})(A\bar{k})(\bar{A}k)(Bk)(\bar{B}\bar{k})] \nn \\
&&  -{1\over 16}{(k\bar{k} + kp - 1)(k\bar{k} - 1)\over kp (kp
-1)}[(AB)(\bar{A}k)(Bk)(\bar{B}\bar{k})(\bar{B}\bar{k}) +
(\bar{A}\bar{B})(A\bar{k})(Bk)(Bk)(\bar{B}\bar{k})\nn \\
&& - (A\bar{A})(Bk)(Bk)(\bar{B}\bar{k})(\bar{B}\bar{k}) -
(k\bar{k})(AB)(\bar{A}\bar{B})(Bk)(\bar{B}\bar{k})] \nn \\
&& - {1\over 16}{k\bar{k} - 1\over
kp}[(A\bar{A})(B\bar{k})(Bk)(\bar{B}k)(\bar{B}\bar{k}) +
(k\bar{k})(A\bar{B})(\bar{A}B)(Bk)(\bar{B}\bar{k}) + \nn \\
&&  - (A\bar{B})(\bar{A}k)(\bar{A}k)(Bk)(B\bar{k})(\bar{B}\bar{k}) -
(A\bar{k})(\bar{A}B)(Bk)(\bar{B}k)(\bar{B}\bar{k})\}\cB(kp +1 , k\bar{k} - 1)
\eea


\newpage

\begin{thebibliography}{99}

\bibitem {ZNS2}
H.~C.~Kao and J.~C.~Lee,
Phys.\ Rev.\ D \textbf{67}, 086003 (2003).
C.~T.~Chan, J.~C.~Lee and Y.~Yang,
Phys.\ Rev.\ D \textbf{71}, 086005 (2005)

 \bibitem{Bianchi:2006nf}
  M.~Bianchi and A.~V.~Santini,
  JHEP {\bf 0612} (2006) 010
  [arXiv:hep-th/0607224].

\bibitem{Ko:2008ft}
  S.~L.~Ko, J.~C.~Lee and Y.~Yang,
  JHEP {\bf 0906} (2009) 028
  [arXiv:0812.4190 [hep-th]].

\bibitem{Feng:2010yx}
  W.~Z.~Feng, D.~Lust, O.~Schlotterer, S.~Stieberger and T.~R.~Taylor,
  Nucl.\ Phys.\  B {\bf 843} (2011) 570
  [arXiv:1007.5254 [hep-th]].

\bibitem{Bianchi:2010es}
  M.~Bianchi, L.~Lopez and R.~Richter,
  JHEP {\bf 1103} (2011) 051
  [arXiv:1010.1177 [hep-th]].

\bibitem{Schlotterer:2010kk}
  O.~Schlotterer,
  Nucl.\ Phys.\  B {\bf 849} (2011) 433
  [arXiv:1011.1235 [hep-th]].


\bibitem{Bianchi:2010dy}
  M.~Bianchi and L.~Lopez,
  JHEP {\bf 1007} (2010) 065
  [arXiv:1002.3058 [hep-th]].


\bibitem{TesiLucaL} { Luca Lopez},\,{ Tesi di laurea specialistica}

\bibitem{Maldacena:1997re}
  J.~M.~Maldacena,
  Adv.\ Theor.\ Math.\ Phys.\  {\bf 2} (1998) 231
  [Int.\ J.\ Theor.\ Phys.\  {\bf 38} (1999) 1113]
  [arXiv:hep-th/9711200].


\bibitem{Gubser:1998bc}
  S.~S.~Gubser, I.~R.~Klebanov and A.~M.~Polyakov,
  Phys.\ Lett.\  B {\bf 428} (1998) 105
  [arXiv:hep-th/9802109].




\bibitem{Witten:1998qj}
  E.~Witten,
  Adv.\ Theor.\ Math.\ Phys.\  {\bf 2} (1998) 253
  [arXiv:hep-th/9802150].



\bibitem{Aharony:1999ti}
  O.~Aharony, S.~S.~Gubser, J.~M.~Maldacena, H.~Ooguri and Y.~Oz,
  Phys.\ Rept.\  {\bf 323} (2000) 183
  [arXiv:hep-th/9905111].



\bibitem{Lee:2011ita}
  J.~C.~Lee, Y.~Mitsuka and Y.~Yang,
and
  arXiv:1101.1228 [hep-th].


\bibitem{D'Appollonio:2010ae}
  G.~D'Appollonio, P.~Di Vecchia, R.~Russo and G.~Veneziano,
  JHEP {\bf 1011} (2010) 100
  [arXiv:1008.4773 [hep-th]].



\bibitem{Green:1987mn}
  M.~B.~Green, J.~H.~Schwarz and E.~Witten,
{\it  Cambridge, Uk: Univ. Pr. ( 1987) 596 P. ( Cambridge Monographs On
Mathematical Physics)}


\bibitem{Green:1987sp}
  M.~B.~Green, J.~H.~Schwarz and E.~Witten,
{\it  Cambridge, Uk: Univ. Pr. ( 1987) 469 P. ( Cambridge Monographs On
Mathematical Physics)}



\bibitem{Gross:1987kza}
  D.~J.~Gross and P.~F.~Mende,
  Phys.\ Lett.\  B {\bf 197}, 129 (1987).



\bibitem{Gross:1987ar}
  D.~J.~Gross and P.~F.~Mende,
  Nucl.\ Phys.\  B {\bf 303}, 407 (1988).


\bibitem{Amati:1987wq}
  D.~Amati, M.~Ciafaloni and G.~Veneziano,
  Phys.\ Lett.\  B {\bf 197}, 81 (1987).

  \bibitem{Amati:1987uf}
  D.~Amati, M.~Ciafaloni and G.~Veneziano,
  Int.\ J.\ Mod.\ Phys.\  A {\bf 3}, 1615 (1988).

\bibitem{Amati:1988tn}
  D.~Amati, M.~Ciafaloni and G.~Veneziano,
  Phys.\ Lett.\  B {\bf 216}, 41 (1989).

  \bibitem{Amati:1990xe}
  D.~Amati, M.~Ciafaloni and G.~Veneziano,
PLANCKIAN
  Nucl.\ Phys.\  B {\bf 347}, 550 (1990).


\bibitem{Amati:1993tb}
  D.~Amati, M.~Ciafaloni and G.~Veneziano,
  Nucl.\ Phys.\  B {\bf 403}, 707 (1993).

\bibitem{Amati:1992zb}
  D.~Amati, M.~Ciafaloni and G.~Veneziano,
  Phys.\ Lett.\  B {\bf 289}, 87 (1992).



\bibitem{Black:2011ep}
  W.~Black and C.~Monni,
  arXiv:1107.4321 [hep-th].

\bibitem{Fotopoulos:2010cm}
  A.~Fotopoulos, N.~Prezas,
  Nucl.\ Phys.\  {\bf B845 } (2011)  340-380.
  [arXiv:1009.3903 [hep-th]].

\bibitem{Friedan:1985ge}
  D.~Friedan, E.~J.~Martinec and S.~H.~Shenker,
  Nucl.\ Phys.\  B {\bf 271} (1986) 93.


\bibitem{Skliros:2011si}
  D.~Skliros, M.~Hindmarsh,
[arXiv:1107.0730 [hep-th]].

\bibitem{Hanany:2010da}
  A.~Hanany, D.~Forcella and J.~Troost,
  Nucl.\ Phys.\  B {\bf 846} (2011) 212
  [arXiv:1007.2622 [hep-th]].

\bibitem{Beisert:2004di}
  N.~Beisert, M.~Bianchi, J.~F.~Morales and H.~Samtleben,
  JHEP {\bf 0407} (2004) 058
  [arXiv:hep-th/0405057].

\bibitem{Bianchi:2003wx}
  M.~Bianchi, J.~F.~Morales and H.~Samtleben,
  JHEP {\bf 0307} (2003) 062
  [arXiv:hep-th/0305052].

\bibitem{Bianchi:2004xi}
  M.~Bianchi,
  Fortsch.\ Phys.\  {\bf 53} (2005) 665
  [arXiv:hep-th/0409304].

 \bibitem{Garousi:1996ad}
  M.~R.~Garousi and R.~C.~Myers,
  Nucl.\ Phys.\  B {\bf 475} (1996) 193
  [arXiv:hep-th/9603194].

\bibitem{Hashimoto:1996bf}
  A.~Hashimoto and I.~R.~Klebanov,
  Nucl.\ Phys.\ Proc.\ Suppl.\  {\bf 55B}, 118 (1997)
  [arXiv:hep-th/9611214].

\bibitem{Becker:2011bw}
  K.~Becker, G.~Guo, D.~Robbins,
  [arXiv:1106.3307 [hep-th]].

\bibitem{Gubser:1997qr}
  S.~S.~Gubser,
  Phys.\ Rev.\  D {\bf 56} (1997) 4984
  [arXiv:hep-th/9704195].

\bibitem {Callan:1996tv}
  C.~G.~.~Callan, S.~S.~Gubser, I.~R.~Klebanov and A.~A.~Tseytlin,
  Nucl.\ Phys.\  B {\bf 489} (1997) 65
  [arXiv:hep-th/9610172].

\end{thebibliography}


\end{document}